# A Comparative Review of RNA Language Models

He Wang[1], Yikun Zhang[1,2], Jie Chen[2], Jian Zhan[1,3,*], Yaoqi Zhou[1,*]

[1] Institute of Systems and Physical Biology, Shenzhen Bay Laboratory, Shenzhen 518107, China
[2] School of Electronic and Computer Engineering, Peking University, Shenzhen 518055, China
[3] Ribopeutic Inc, Guangzhou International Bio Island, Guangdong 510320, China

* To whom correspondence should be addressed. Tel: +86 755 26849275; Email: zhouyq@szbl.ac.cn

Correspondence may also be addressed to Jian Zhan. Tel: +86 755 26849280; Email: zhanjian@szbl.ac.cn

## Abstract

Given usefulness of protein language models (LMs) in structure and functional inference, RNA LMs have received increased attentions in the last few years. However, these RNA models are often not compared against the same standard. Here, we divided RNA LMs into three classes (pretrained on multiple RNA types (especially noncoding RNAs), specific-purpose RNAs, and LMs that unify RNA with DNA or proteins or both) and compared 13 RNA LMs along with 3 DNA and 1 protein LMs as controls in zero-shot prediction of RNA secondary structure and functional classification. Results shows that the models doing well on secondary structure prediction often perform worse in function classification or vice versa, suggesting that more balanced unsupervised training is needed.

## Introduction

Recent advances in language models (LMs) have revolutionized the ability of deep learning techniques in understanding, parsing, and generating natural language by capturing the permutations and relationships between words[1–7]. One of the most iconic and meta-models is Transformer[8], an encoder-decoder architecture that captures dependency among sequence tokens through the self-attention mechanism. Examples of transformer-based models are T5 (Text-to-Text Transfer Transformer)[7], BERT (Bidirectional Encoder Representations from Transformers)[2], and GPT (Generative Pre-trained Transformer)[3,4]. These increasingly accurate LMs begin to blur the boundary between human and artificial intelligence.

Much like 26-letter-coded English language, biopolymers like DNA, RNA, and proteins were also coded by a few letters (typically 4 for DNA and RNA, and 20 for proteins). As a result, advances in natural language models have also led to progress in protein, RNA, and DNA LMs.

Various protein LMs were developed including T5-based ProtT5[9], GPT-based ProGPT2[10] and BERT-based ProteinBert[11] and evolutionary scale modeling (ESM)[12]. In addition to single-sequence-based models, ESM-MSA-1b (or known as MSA Transformer)[13] was also developed based on multiple sequence alignment (MSA). Compared to GPT structure, BERT uses a bidirectional self-attention mechanism, which means that it can focus on both left and right-side information of the input sequence instead of merely using the previous information to infer the later output. Moreover, Lin et al. showed that the unsupervised contact results of ESM-2 (BERT-style model) are better than ProtT5 (encoder-decoder architecture) for 3B model parameters[14]. As a result, BERT is more often employed as a generalized foundation model for biological sequence studies[12]. The rapid development of protein LMs has driven prediction and design of protein structure and function[12,15], included as a core module in protein 3D structure prediction models such as ESMfold[14].

Following the protein LMs, various DNA LMs and RNA LMs have been proposed. One of the representative DNA foundational models was DNABERT[16]. It adopted a BERT-style structure pre-trained on the human genome sequence represented as $k$-mers tokens, outperforming previous Convolutional Neural Network (CNN)-based and Recurrent Neural Network (RNN)-based models in the prediction of promoters, splice sites, and transcription factor binding sites. RNA-FM[17] was the first released (but not yet published) RNA foundation model pre-trained on whole-scale unannotated ncRNAs. This BERT-style model was soon followed by other BERT models with different model sizes and training sets (Table 1) along with various applications, including RNA classification[17,18], secondary structure prediction[19,20], and mean ribosome load (MRL) prediction[21]. Only one exception is an MSA-transformer-based model called RNA-MSM[20]. In addition, the unified language model capable of processing RNA as well as DNA or proteins, e.g. LucaOne[22], has been proposed and can be also used to study RNA. In this survey, we focus on evaluating the performance of existing RNA LMs on ncRNAs and compare them with representative protein LM and DNA LM as "controls".

## Results

## Current RNA language model types and BERT extensions

Table 1 lists RNA LMs published along with some found in preprint servers. These models can be separated into three classes: those pre-trained on nonspecific various types of RNAs, especially noncoding RNAs (Class I) [17–20,23–26], those pre-trained on specific datasets for specific purposes (Class II) [21,27] and those pre-trained or tested on not only RNA datasets but also DNA or proteins datasets (Class III: unified LMs)[22,28]. Class I models are often used as RNA fundamental models, enabling generalization to study the structural and functional properties of a wide range of RNAs. Most of them are pre-trained on the whole or nonredundant subsets of the RNAcentral database, a comprehensive collection of all ncRNA types from a broad range of organisms (> 32M ncRNA sequences for the latest version)[29]. One exception is MP-RNA, which was pre-trained on the dataset consisting of mRNA, CDS and UTRs[30]. The second is MSA-Transformer-based RNA-MSM[20], which is limited to multiple sequence alignment generated by RNAcmap3[31] for ~4000 RNA families from Rfam[32], largely due to limited computational resource for generating homologous sequences. The third is UNI-RNA[18] pretrained on the Master database of all possible RNA Sequences (MARS), which include RNAcentral, the transcriptome assembly, metagenome assembly and genomic sequences (1B sequences)[31]. The Class II models, by contrast, employed a specific type of RNA (e.g. 5'UTR[21] and pre-mRNA[27]) for pretraining. The Class I and II RNA LMs differ in hyperparameters employed with model sizes ranging from 1M in UTR_LM[21] to 1.6B in AIDO.RNA[33], all of which are far smaller than 15B parameters utilized in protein LM ESM2[14]. The Class III models, also known as unified LMs, are not purely RNA LMs, but can be used as a foundation model for RNA[22].

As Table 1 shows, most RNA LMs are based on a BERT-style model. The basic network for most LMs consisting of three main steps (Fig. 1): the input layer (including masking and embedding), the encoding layer, and the output processing[2]. First, the input RNA sequence is segmented and converted into a series of tokens, e.g., four bases ("A", "U", "G", and "C"). Next, during the masking process, typically 15% of the tokens from each training sequence were randomly selected. Of these, 80% are masked, i.e., replaced with "[mask]" tokens, 10% are replaced with randomly selected tokens from the vocabulary, and the remaining 10% remain unchanged. Then, these tokens are converted into word vectors and fed into the encoding layer, which is the core part of the BERT model and consists of stacked multiple Transformer encoders. The whole encoding process can be viewed as a special aggregation operation on the input sequence to obtain a more comprehensive and richer representation. Finally, the output processing is determined by the masked language model (MLM)[2], which requires the BERT to predict the original form of those masked tokens, prompting the model to learn the ability to make predictions with incomplete contextual information. The training process of the RNA LMs usually consists of two stages: pre-training and fine-tuning. The pre-training stage typically requires large-scale RNA sequences to ensure that the model learns the general semantic representation. Attention maps and output sequence embeddings of the pretrained model can be extracted for downstream RNA structure and function prediction. In the fine-tuning stage, the foundation model can be further fine-tuned according to specific datasets to obtain better performance than original model.

Some models extend the above BERT-like model training with new strategies. Because the key feature for RNA secondary structure is canonical Watson-Crick (AU/GC) and wobble (GU) base pairing, ERNIE-RNA[19] introduced pairwise-position-based pairing scores biased toward AU, GC and GU, as well as non-canonical base pairs. RNAErnie employed pre-annotated RNA sequence motifs for motif-level random masking at base/subsequence levels[25]. RNA-km incorporated $k$-mers (where $3 \leq k \leq 8$) rather than single-base masking strategy to enhance capture of local dependencies[26]. ProtRNA extended the vocabulary, embedding layer, and bias layer of the original protein model ESM2 to accommodate added RNA tokens, and adapted the pre-trained ESM-2 on protein to ncRNA via transfer learning[23]. BiRNA-BERT implemented an adaptive tokenization with Byte Pair Encoding (BPE) tokenization for long sequences and NUC (nucleotide-level) tokenization for shorter ones through dual pretraining strategy[34]. Later, we will refer to BiRNA-BERT using different tokenization as "BiRNA-BERT (BPE)" and "BiRNA-BERT (NUC)", respectively. To consider the effectiveness of secondary structure for the genomic task, MP-RNA integrates secondary structure annotation and single-nucleotide mutation repair into the pre-training in addition to applying the MLM task[30]. Finally, to incorporate evolutionary information directly, RNA-MSM employed MSA-Transformer structure and utilized a set of homologous sequences generated from RNAcmap[31,35] instead of a single input sequence[20].

## The problems associated with model comparison

Most models were evaluated according to downstream tasks, which can be broadly categorized into prediction of structural and functional properties. Table 2 lists structural and functional properties commonly predicted by various RNA models. Structural properties include solvent accessibility prediction[20], secondary structure prediction[17–20,23–26,30,33,34], 3D contact prediction[17,19], and 3D torsion angle prediction[34]. Functional properties include RNA classifications[17,36], splice sites[27], modification sites[18,33], RNA-protein/RNA interactions[37,38], mean ribosome load (MRL)[17–19,21,23,24,33,39], mRNA expression level (EL)[21,33,40], mRNA translation efficiency (TE)[21,33,40], and unannotated internal ribosome entry sites (IRESs) identification[21,39,41].

Often, newer models were compared to previous methods by using the same datasets for downstream tasks. However, there is always a question of about the quality of the test set and potential over-fitting during fine tuning. For example, dominance of rRNA and tRNA is observed in the pre-training dataset RNAcentral used by most RNA LMs, commonly used datasets for downstream RNA secondary structure prediction (RNAStralign[42], ArchiveII[43], and bpRNA[44]), and 20M Rfam sequences. Most ncRNAs are not in 4178 families of Rfam (Table 3). Moreover, some training and test sets did not perform redundancy removal.

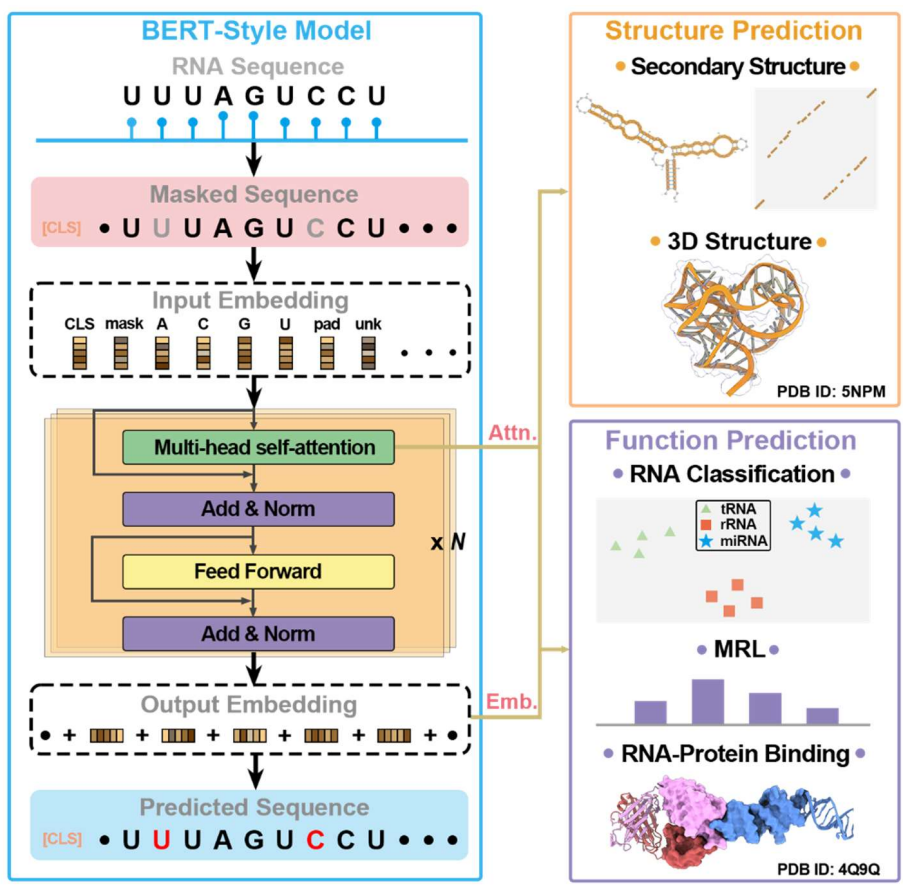

**Fig. 1 | The adoption of BERT-style model for RNA language models.**

## Zero-shot model evaluation: Secondary structure

To have a more consistent quantitative comparison between different language models, we perform the comparison before fine tuning for downstream tasks. That is, we investigate if the model has learned structural or functional properties prior to fine tuning. Previously, we have showed that the attention maps of the 2D form in the RNA-MSM model contain information about the secondary structure of RNA[20] with base-pairing thresholds determined by a 40-RNA validated set (VL1) and evaluated by a 70-RNA test set (TS), whereas embedding layers contains the information on solvent accessibility[20], all without training. To make the evaluation consistent, the best head-layer position for secondary structure found in VL1 is employed to evaluate in TS for all RNAs. We also employed DNA LM (DNABERT) and protein LM (ESM2) as the references for comparison. Although Evo, based on StripedHyena architecture, is also a representative unified LM for analyzing both RNA and DNA, it is not compared with RNA LMs in this survey because Evo's official model does not provide the corresponding code to extract embedding and attentions, nor can these be extracted in a manner like Transformer[28]. For the protein LM ESM2, 30 PDB structures from CASP14 and 29 PDB structures from CASP15[45] were employed as the validation and test sets, respectively. A residue-residue contact is defined as any two amino acids with a "CB" ("CA" for glycine) distance less than 8 Å[46]. All RNA and protein sequences used in this comparison are shorter than or equal to 510 due to sequence length restrictions imposed by some of the LMs.

Table 4 (and Fig. 2a) compares the performance of zero-shot prediction in RNA secondary structure (or protein contact maps) in term of median, mean, and std values for individual RNAs (proteins) using the same head-layer position for all RNAs (proteins). As expected, DNA LMs and Class II RNA LMs contain no secondary-structural information, indicating the importance of ncRNA sequence libraries employed for pre-training of RNA foundation models. Although LucaOne demonstrates the ability to predict secondary structure, it performs less well than Class I RNA LMs, possibly suggesting the limit of a universal model. Among Class I RNA LMs, RNA-MSM has the highest F1 scores, with a median value 9.0% and a mean value 14.6% higher than AIDO.RNA (1.6B), despite its t raining with a small number of families (<4000). This result highlights the difficulty to obtain RNA secondary structure information from single sequence directly, compared to single-sequence LMs of proteins and RNAs with aligned homologs. At the median level, RiNALMo with a model size of 651M improves over RiNALMo with a model size of 148M by 28.3% and 1.6B-AIDO.RNA is 32.2% higher than 648M-AIDO.RNA, highlighting the importance of a larger language model. Unlike proteins with a narrow distribution of F1 scores with the high median value even at a single sequence level (protein ESM2 in Fig. 2a), the performance by BERT-style RNA LMs has a large fluctuation from RNA to RNA, almost completely unpredictable for some RNAs, indicating the challenging facing RNA secondary structure prediction for some RNAs.

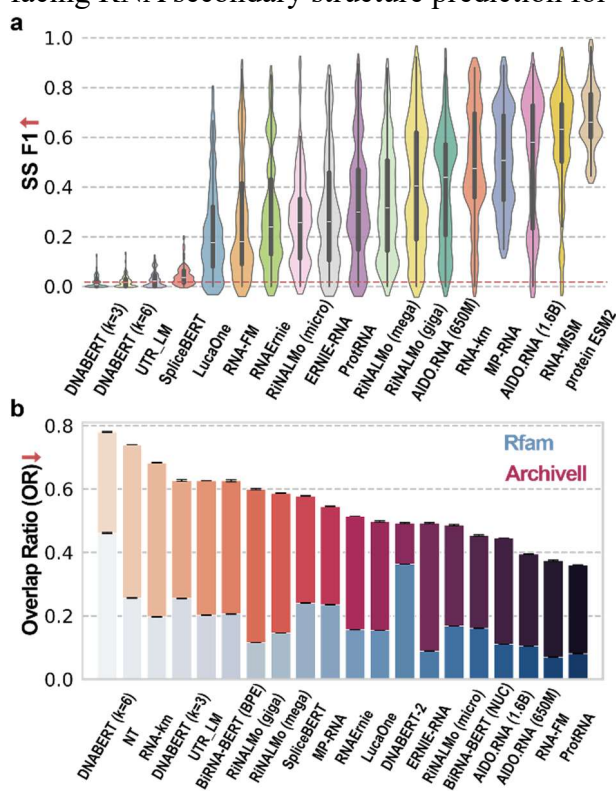

**Fig. 2 | Zero-shot model evaluation for secondary structure prediction and RNA**

**classification. a.** Distribution of secondary structure (SS) F1 scores (harmonic mean of sensitivity and precision) for 70 RNAs in TS, based on the attention map on the optimal header positions evaluated on VL1. As "controls", DNA LMs and protein LM ESM2 (zero-shot protein contact predictions on the CASP15 dataset) are also shown. The higher the F1 score, the better the model. **b.** The overlap ratios (ORs) between the cosine similarity distributions of homologous and non-homologous sequences in the Rfam (white-to-blue gradient) and ArchiveII (red-to-purple gradient) datasets, respectively. A smaller OR value indicates a higher classification performance.

## Zero-shot model evaluation: RNA classifications

To evaluate the ability of the output embeddings of the LMs to discriminate between homologous and non-homologous sequences, 100,000 homologous and 100,000 non-homologous sequence pairs were randomly sampled from the Rfam dataset, where two sequences in a homologous sequence pair have the same RNA family and those in a non-homologous sequence pair have different RNA families. The embeddings of all sampled sequences were extracted from the LMs, and the cosine similarity between two sequence embeddings in each homologous or non-homologous sequence pair was calculated. We employed two datasets for zero-shot RNA classifications: Rfam and ArchiveII (see Methods), and compared RNA-LMs with three representative DNA LMs (DNABERT, DNABERT-2[47] and NT[48]) as controls.

The cosine similarity distributions were shown in Extended Data Fig. 1. Overall, both RNA LMs and DNA LMs can distinguish between homologous and non-homologous distributions, with the cosine similarity in the homologous distribution favoring larger values. This demonstrates that LMs can learn RNA family properties from genes or RNA sequences in an unsupervised manner. To quantify the ability to discriminate between homologous and non-homologous sequences, the overlap ratio (OR) was introduced (see Methods). A smaller OR corresponds to a lower degree of overlap between the homologous and non-homologous distribution curves, indicating that it is easier for the LM to distinguish whether the two sequences are homologous or not. When the two distributions are completely separated, the OR value is equal to 0; conversely, when the two distributions completely overlap, the OR value is equal to 1.

In Table 4 (and Fig. 2b), most Class I RNA LMs generally outperform DNA LMs and Class II RNA LMs on the Rfam dataset, and the performance of the unified LM, LucaOne, corresponds to the mid-range level of Class I RNA LMs. RNA-FM performs best here with OR=0.070±0.000. However, RNA-km and MP-RNA, which perform among the best in secondary structure prediction (behind the MSA-based method RNA-MSM and 1.6B parameter AIDO.RNA only), perform poorly here, at a similar level to Class II RNA LMs. This implies that enhancement in secondary structure may lead to poorer embedding. AIDO.RNA is the only model with high performance both in secondary structure and OR. However, AIDO.RNA (1.6B) improves over AIDO.RNA

(650M) in secondary structure but is worse in OR.

We further did the same comparison for the ArchiveII dataset, except that here sequences with the same RNA type were treated as homologous and those with different RNA types were treated as non-homologous. The cosine similarity distributions were shown in Extended Data Fig. 2. Although Rfam families appear to be much easier to separate than the families in the ArchiveII dataset, similar pattern was found with a strong Pearson correlation coefficient of 0.68 between the results on Rfam and ArchiveII datasets. (Some models that perform well on the Rfam, however, may perform poorly on the more difficult ArchiveII, e.g. RiNALMo.) Rfam families were easier to separate than ArchiveII because the sequence identities within Rfam families (an average of 77.1%) are much higher than the sequence identities within ArchiveII families (an average of 54.5%).

## Discussion

In the past decade, research and clinical development based on the physiological functions of RNAs, especially the ncRNAs, have been on the rise[49–52]. It is well known that structure is the basis of function. However, the currently resolved RNA structures are fewer than 0.1% of known sequences. The success of protein LMs in structure and function studies for proteins led to a surge interest of RNA LMs for investigating RNAs [17,18,20,24]. To compare these LMs on the same standard, we performed zero-shot prediction of secondary structure and functional classification with the following observations.

First, current prediction of RNA secondary structure from an RNA LM has a lower F1 score than prediction of protein contact maps. This is perhaps due to much smaller model sizes in RNA LMs (1.6B-AIDO.RNA improved secondary structure prediction by 32.2% over 648M-AIDO.RNA, current largest 15B-ESM2 model is ~9 times larger than the 1.6B-AIDO.RNA model). However, F1-score of protein ESM2 models is nearly independent of model sizes (the mean F1 scores are 0.672, 0.663, 0.675, 0.677, 0.623, 0.636 for 8M, 35M, 150M, 650M, 3B, 15B ESM2 models, respectively). Thus, this is more likely because RNA sequences are not as conserved as proteins due to a much smaller number of letter codes[17,20,53]. As a result, it is more difficult to recognize correlated mutations. Indeed, MSA-based RNA-MSM provides a substantially improved prediction of RNA secondary structures despite trained on a very small dataset. Second, some models (e.g. RNA-km and MP-RNA) performed exceptionally well on secondary structure but extremely poor in RNA classification, suggesting unbalanced training. Third, unified LM models are poor in RNA-specific prediction. We found that LucaOne is typically worse than the Class I models in RNA secondary structure prediction and RNA classification.

Thus, there are significant rooms for further improvement of RNA LMs regarding how to make an unbiased training in structure and function. Larger model sizes will surely help. However, it is not certain if a large RNA database (MARS) will also help because UNI-RNA is not yet openly available. More importantly, how to improve secondary structure prediction without hurting RNA classification deserves special attentions to overcome the poor performance revealed from comparing 1.6B-AIDO.RNA model against 648M-AIDO.RNA and the results of RNA-km and MP-RNA.

## Methods

**Dataset for zero-shot RNA secondary structure prediction.** The validation dataset (VL1) used for threshold assessment included 40 RNAs with PDB experimental structures, and the test dataset (TS) contained 70 RNAs with PDB experimental structures[20]. Moreover, to test the zero-shot contact prediction of protein LMs, 30 PDB structures from CASP14 and 29 PDB structures from CASP15 were employed as the validation and test sets, respectively[45]. These sequences are all less than 510 in length.

**Dataset for zero-shot RNA classification experiment**. Three sets of ncRNA sequence datasets were prepared. The first is labeled as Rfam, which contains 24,607 sequences sampled from 4170 families in Rfam 14.10 seed alignments, with up to 10 sequences randomly selected from each RNA family. Of these, there are 238 sequences over 510 nt and 84 sequences over 1022 nt. The second is ArchiveII, which contains 3,975 sequences that can be categorized into 10 RNA types. Since the maximum sequence length allowed by some of the models is 510 (excluding the append and prepend markers), and in order to be consistent with previous results, only sequences with lengths less than or equal to 510 were taken here, totaling 3,864 sequences containing 9 RNA types.

**Zero-shot secondary structure prediction**. First, RNA sequences in VL1 and TS were input into the language model and the attention maps at each head-layer position were extracted. Each head-layer attention map was transformed into the secondary structure probabilities after symmetrization, APC correction and sigmoid function processing[13]. If the probability is greater than the preset threshold, the corresponding two bases would be predicted to be paired. We next set thresholds within the range [0, 1] at intervals of 0.001 and computed the F1 score for secondary structure predictions from each attention map on VL1. And the optimal head-layer position and threshold were selected. Finally, with this head-layer position and threshold, the F1 score for secondary structure prediction on TS were evaluated. A similar process is used for zero-shot protein contact prediction, with the difference that the transition from attention maps to secondary structure probabilities in RNA is treated as residue-residue contact probabilities in proteins.

**Calculation of the cosine similarity between two sequence embeddings**. The embeddings of RNA sequence A and B were first extracted from the language model, and dimensionally transformed with the Fourier Transform (FFT) method that converts the embedding of $x \in \mathbb{R}^{L \times d}$ into a representational vector of $x' \in \mathbb{R}^{d'}$, where $L$ is the sequence length, $d$ is the embedding dimension, and $d'$ is the dimension of the representational vector that is set 128 here. Then, the cosine similarity is computed as follows:

$$cs = \frac{a \cdot b}{|a||b|}$$

Where $a$ and $b$ were the representational vectors of sequence A and B, respectively; $|x|$ denotes the modulo operation; $cs$ was the cosine similarity and has the value in the range $[0, 1]$.

**Calculation of the overlap ratio (OR)**. OR is defined as the area of intersection of two distribution curves under the same coordinate system divided by the area of their concatenation, and its value is distributed in [0, 1]. The larger the OR value, the more the two distribution curves overlap. If the OR value is equal to 0, it means that the two distributions are completely separated, and if it is equal to 1, it means that the two distributions completely overlap. In the RNA classification experiments, the OR between the cosine similarity distribution of sampled homologous sequences and the cosine similarity distribution of sampled non-homologous sequences was calculated, and the average of the ORs under three independent random samples was performed.

# Data Availability

Datasets for zero-shot RNA classifications and RNA secondary structure prediction as well as protein contact map predictions are available at https://zenodo.org/records/14430869.

# Code Availability

Codes and datasets for zero-shot RNA classifications and RNA secondary structure prediction as well as protein contact map predictions are available at https://zenodo.org/records/14430869. RNA-MSM was downloaded via GitHub at https://github.com/yikunpku/RNA-MSM (2024/10/8); SpliceBERT was downloaded via GitHub at https://github.com/chenkenbio/SpliceBERT (2024/8/11); UTR_LM was downloaded via GitHub at https://github.com/a96123155/UTR-LM (2024/8/12); RNA-FM was downloaded via GitHub at https://github.com/ml4bio/RNA-FM (2024/7/26); RiNALMo was downloaded via GitHub at https://github.com/lbcb-sci/RiNALMo (2024/7/26); ERNIE-RNA was downloaded via GitHub at https://github.com/Bruce-ywj/ERNIE-RNA (2024/7/26); RNAErnie was downloaded via GitHub at https://github.com/CatIIIIIIII/RNAErnie (2024/7/26); RNA-km was downloaded via GitHub at https://github.com/gongtiansu/RNA-km (2024/9/6); ProtRNA was downloaded via GitHub at https://github.com/roxie-zhang/ProtRNA (2024/9/24); DNABERT was downloaded via GitHub at https://github.com/jerryji1993/DNABERT (2024/9/27); ESM2 (150M) was downloaded via GitHub at

https://github.com/facebookresearch/esm (2024/4/12); DNlABERT-2 was downloaded via GitHub at https://github.com/MAGICS-LAB/DNABERT_2 (2024/8/13); NT was downloaded via GitHub at https://github.com/instadeepai/nucleotide-transformer (2024/8/26); BiRNA-BERT was downloaded via GitHub at https://github.com/buetnlpbio/BiRNA-BERT (2024/10/8); LucaOne was downloaded via GitHub at https://github.com/LucaOne/LucaOne (2024/11/25); AIDO.RNA was downloaded via GitHub at https://github.com/genbio-ai/AIDO (2024/12/11); MP-RNA was downloaded via Hugging Face at https://huggingface.co/yangheng/MP-RNA (2024/11/27).

## Acknowledgements

We acknowledge the support of the Shenzhen Bay supercomputing facility and Pengcheng Cloudbrain. This work was supported by the Natural Science Foundation of China [22350710182 to Y.Z.], the Shenzhen Medical Research Funds in China (No. B2302037 to J.C.), and the National Key R&D Program of China [2022ZD0118201 to J.C.].

## Reference

1. Zhang, Q. *et al.* Scientific Large Language Models: A Survey on Biological & Chemical Domains. Preprint at http://arxiv.org/abs/2401.14656 (2024).
2. Devlin, J., Chang, M.-W., Lee, K. & Toutanova, K. BERT: Pre-training of Deep Bidirectional Transformers for Language Understanding. *ArXiv E-Prints* arXiv:1810.04805 (2018) doi:10.48550/arXiv.1810.04805.
3. Radford, A. & Narasimhan, K. Improving Language Understanding by Generative Pre-Training. in (2018).
4. Brown, T. B. *et al.* Language Models are Few-Shot Learners. *ArXiv E-Prints* arXiv:2005.14165 (2020) doi:10.48550/arXiv.2005.14165.
5. Touvron, H. *et al.* LLaMA: Open and Efficient Foundation Language Models. *ArXiv E-Prints* arXiv:2302.13971 (2023) doi:10.48550/arXiv.2302.13971.
6. Chowdhery, A. *et al.* PaLM: scaling language modeling with pathways. *J Mach Learn Res* **24**, (2024).
7. Raffel, C. *et al.* Exploring the Limits of Transfer Learning with a Unified Text-to-Text Transformer. *J. Mach. Learn. Res.* **21**, 1–67 (2020).
8. Vaswani, A. *et al.* Attention Is All You Need. *ArXiv E-Prints* arXiv:1706.03762 (2017) doi:10.48550/arXiv.1706.03762.
9. Elnaggar, A. *et al.* ProtTrans: Toward Understanding the Language of Life Through Self-Supervised Learning. *IEEE Trans. Pattern Anal. Mach. Intell.* **44**, 7112–7127 (2022).
10. Ferruz, N., Schmidt, S. & Höcker, B. ProtGPT2 is a deep unsupervised language model for protein design. *Nat. Commun.* **13**, 4348 (2022).
11. Brandes, N., Ofer, D., Peleg, Y., Rappoport, N. & Linial, M. ProteinBERT: a universal deep-learning model of protein sequence and function. *Bioinforma. Oxf. Engl.* **38**, 2102–2110 (2022).

12. Rives, A. *et al.* Biological structure and function emerge from scaling unsupervised learning to 250 million protein sequences. *Proc. Natl. Acad. Sci. U. S. A.* **118**, (2021).
13. Rao, R. *et al.* MSA Transformer. *bioRxiv* (2021) doi:10.1101/2021.02.12.430858.
14. Lin, Z. *et al.* Evolutionary-scale prediction of atomic-level protein structure with a language model. *Science* **379**, 1123–1130 (2023).
15. Rao, R. *et al.* Evaluating Protein Transfer Learning with TAPE. Preprint at https://doi.org/10.1101/676825 (2019).
16. Ji, Y., Zhou, Z., Liu, H. & Davuluri, R. V. DNABERT: pre-trained Bidirectional Encoder Representations from Transformers model for DNA-language in genome. *Bioinforma. Oxf. Engl.* **37**, 2112–2120 (2021).
17. Chen, J. *et al.* Interpretable RNA Foundation Model from Unannotated Data for Highly Accurate RNA Structure and Function Predictions. (2022).
18. Wang, X. *et al.* UNI-RNA: UNIVERSAL PRE-TRAINED MODELS REVOLUTIONIZE RNA RESEARCH. Preprint at https://doi.org/10.1101/2023.07.11.548588 (2023).
19. Yin, W. *et al.* ERNIE-RNA: An RNA Language Model with Structure-enhanced Representations. Preprint at https://doi.org/10.1101/2024.03.17.585376 (2024).
20. Zhang, Y. *et al.* Multiple sequence alignment-based RNA language model and its application to structural inference. *Nucleic Acids Res.* **52**, e3–e3 (2024).
21. Chu, Y. *et al.* A 5′ UTR language model for decoding untranslated regions of mRNA and function predictions. *Nat. Mach. Intell.* **6**, 449–460 (2024).
22. He, Y. *et al.* LucaOne: Generalized Biological Foundation Model with Unified Nucleic Acid and Protein Language. Preprint at https://doi.org/10.1101/2024.05.10.592927 (2024).
23. Zhang, R., Ma, B., Xu, G. & Ma, J. ProtRNA: A Protein-derived RNA Language Model by Cross-Modality Transfer Learning. Preprint at https://doi.org/10.1101/2024.09.10.612218 (2024).
24. Penić, R. J., Vlašić, T., Huber, R. G., Wan, Y. & Šikić, M. RiNALMo: General-purpose RNA language models can generalize well on structure prediction tasks. Preprint at http://arxiv.org/abs/2403.00043 (2024).
25. Wang, N. *et al.* Multi-purpose RNA language modelling with motif-aware pretraining and type-guided fine-tuning. *Nat. Mach. Intell.* **6**, 548–557 (2024).
26. Gong, T. & Bu, D. Language models enable zero-shot prediction of RNA secondary structure including pseudoknots.
27. Chen, K. *et al.* Self-supervised learning on millions of primary RNA sequences from 72 vertebrates improves sequence-based RNA splicing prediction. *Brief. Bioinform.* **25**, bbae163 (2024).
28. Nguyen, E. *et al.* Sequence modeling and design from molecular to genome scale with Evo. *Science* **386**, eado9336 (2024).
29. RNAcentral 2021: secondary structure integration, improved sequence search and new member databases. *Nucleic Acids Res.* **49**, D212–D220 (2021).
30. Yang, H. & Li, K. MP-RNA: Unleashing Multi-species RNA Foundation Model via Calibrated Secondary Structure Prediction.
31. Chen, K., Litfin, T., Singh, J., Zhan, J. & Zhou, Y. MARS and RNAcmap3: The

Master Database of All Possible RNA Sequences Integrated with RNAcmap for RNA Homology Search. *Genomics Proteomics Bioinformatics* **22**, qzae018 (2024).

32. Kalvari, I. *et al.* Rfam 14: expanded coverage of metagenomic, viral and microRNA families. *Nucleic Acids Res.* **49**, D192–D200 (2021).

32. Zou, S. *et al.* A Large-Scale Foundation Model for RNA Function and Structure Prediction. *NeurIPS 2024 Workshop on AI for New Drug Modalities* (2024).

34. Tahmid, M. T., Shahgir, H. S., Mahbub, S., Dong, Y. & Bayzid, Md. S. BiRNA-BERT allows efficient RNA language modeling with adaptive tokenization. Preprint at https://doi.org/10.1101/2024.07.02.601703 (2024).

35. Zhang, T. *et al.* RNAcmap: a fully automatic pipeline for predicting contact maps of RNAs by evolutionary coupling analysis. *Bioinforma. Oxf. Engl.* **37**, 3494–3500 (2021).

36. Rossi, E., Monti, F., Bronstein, M. & Liò, P. ncRNA Classification with Graph Convolutional Networks. Preprint at http://arxiv.org/abs/1905.06515 (2019).

37. Sun, S., Wang, W., Peng, Z. & Yang, J. RNA inter-nucleotide 3D closeness prediction by deep residual neural networks. *Bioinforma. Oxf. Engl.* **37**, 1093–1098 (2021).

38. Xu, Y. *et al.* PrismNet: predicting protein-RNA interaction using in vivo RNA structural information. *Nucleic Acids Res.* **51**, W468–W477 (2023).

39. Sample, P. J. *et al.* Human 5' UTR design and variant effect prediction from a massively parallel translation assay. *Nat. Biotechnol.* **37**, 803–809 (2019).

40. Cao, J. *et al.* High-throughput 5' UTR engineering for enhanced protein production in non-viral gene therapies. *Nat. Commun.* **12**, 4138 (2021).

41. Karollus, A., Avsec, Ž. & Gagneur, J. Predicting mean ribosome load for 5'UTR of any length using deep learning. *PLoS Comput. Biol.* **17**, e1008982 (2021).

42. Tan, Z., Fu, Y., Sharma, G. & Mathews, D. H. TurboFold II: RNA structural alignment and secondary structure prediction informed by multiple homologs. *Nucleic Acids Res.* **45**, 11570–11581 (2017).

43. Sloma, M. F. & Mathews, D. H. Exact calculation of loop formation probability identifies folding motifs in RNA secondary structures. *RNA N. Y. N* **22**, 1808–1818 (2016).

44. Danaee, P. *et al.* bpRNA: large-scale automated annotation and analysis of RNA secondary structure. *Nucleic Acids Res.* **46**, 5381–5394 (2018).

45. Das, R. *et al.* Assessment of three-dimensional RNA structure prediction in CASP15. *Proteins* **91**, 1747–1770 (2023).

46. Singh, J., Litfin, T., Singh, J., Paliwal, K. & Zhou, Y. SPOT-Contact-LM: improving single-sequence-based prediction of protein contact map using a transformer language model. *Bioinformatics* **38**, 1888–1894 (2022).

47. Zhou, Z. *et al.* DNABERT-2: Efficient Foundation Model and Benchmark For Multi-Species Genome. Preprint at http://arxiv.org/abs/2306.15006 (2024).

48. Dalla-Torre, H. *et al.* The Nucleotide Transformer: Building and Evaluating Robust Foundation Models for Human Genomics.

49. Chaudhary, N., Weissman, D. & Whitehead, K. A. mRNA vaccines for infectious diseases: principles, delivery and clinical translation. *Nat. Rev. Drug Discov.* **20**, 817–

838 (2021).

50. Rohner, E., Yang, R., Foo, K. S., Goedel, A. & Chien, K. R. Unlocking the promise of mRNA therapeutics. *Nat. Biotechnol.* **40**, 1586–1600 (2022).

51. Garner, A. L. Contemporary Progress and Opportunities in RNA-Targeted Drug Discovery. *ACS Med. Chem. Lett.* **14**, 251–259 (2023).

52. Childs-Disney, J. L. *et al.* Targeting RNA structures with small molecules. *Nat. Rev. Drug Discov.* **21**, 736–762 (2022).

53. Menzel, P., Gorodkin, J. & Stadler, P. F. The tedious task of finding homologous noncoding RNA genes. *RNA* **15**, 2075–2082 (2009).

54. Sayers, E. W. *et al.* Database resources of the National Center for Biotechnology Information in 2023. *Nucleic Acids Res.* **51**, D29–D38 (2023).

55. Chen, M. *et al.* Genome Warehouse: A Public Repository Housing Genome-Scale Data. *Genomics Proteomics Bioinformatics* **19**, 584–589 (2021).

56. Martin, F. J. *et al.* Ensembl 2023. *Nucleic Acids Res.* **51**, D933–D941 (2023).

57. Leebens-Mack, J. H. *et al.* One thousand plant transcriptomes and the phylogenomics of green plants. *Nature* **574**, 679–685 (2019).

58. Haeussler, M. *et al.* The UCSC Genome Browser database: 2019 update. *Nucleic Acids Res.* **47**, D853–D858 (2019).

59. Cavallo, L., Kleinjung, J. & Fraternali, F. POPS: A fast algorithm for solvent accessible surface areas at atomic and residue level. *Nucleic Acids Res.* **31**, 3364–3366 (2003).

60. Singh, J., Hanson, J., Paliwal, K. & Zhou, Y. RNA secondary structure prediction using an ensemble of two-dimensional deep neural networks and transfer learning. *Nat. Commun.* **10**, 5407 (2019).

61. Singh, J. *et al.* Improved RNA secondary structure and tertiary base-pairing prediction using evolutionary profile, mutational coupling and two-dimensional transfer learning. *Bioinforma. Oxf. Engl.* **37**, 2589–2600 (2021).

62. Leontis, N. B. & Zirbel, C. L. Nonredundant 3D Structure Datasets for RNA Knowledge Extraction and Benchmarking. in *RNA 3D Structure Analysis and Prediction* (eds. Leontis, N. & Westhof, E.) 281–298 (Springer Berlin Heidelberg, Berlin, Heidelberg, 2012). doi:10.1007/978-3-642-25740-7_13.

63. Singh, J., Paliwal, K., Singh, J. & Zhou, Y. RNA Backbone Torsion and Pseudotorsion Angle Prediction Using Dilated Convolutional Neural Networks. *J. Chem. Inf. Model.* **61**, 2610–2622 (2021).

64. Scalzitti, N. *et al.* Spliceator: multi-species splice site prediction using convolutional neural networks. *BMC Bioinformatics* **22**, 561 (2021).

65. Song, Z. *et al.* Attention-based multi-label neural networks for integrated prediction and interpretation of twelve widely occurring RNA modifications. *Nat. Commun.* **12**, 4011 (2021).

66. Han, Y. & Zhang, S.-W. ncRPI-LGAT: Prediction of ncRNA-protein interactions with line graph attention network framework. *Comput. Struct. Biotechnol. J.* **21**, 2286–2295 (2023).

67. Wen, M., Cong, P., Zhang, Z., Lu, H. & Li, T. DeepMirTar: a deep-learning approach for predicting human miRNA targets. *Bioinforma. Oxf. Engl.* **34**, 3781–3787

(2018).

68. Kang, Q., Meng, J., Cui, J., Luan, Y. & Chen, M. PmliPred: a method based on hybrid model and fuzzy decision for plant miRNA–lncRNA interaction prediction. *Bioinformatics* **36**, 2986–2992 (2020).

69. Kolekar, P., Pataskar, A., Kulkarni-Kale, U., Pal, J. & Kulkarni, A. IRESPred: Web Server for Prediction of Cellular and Viral Internal Ribosome Entry Site (IRES). *Sci. Rep.* **6**, 27436 (2016).

70. Weingarten-Gabbay, S. *et al.* Systematic discovery of cap-independent translation sequences in human and viral genomes. *Science* **351**, aad4939 (2016).

71. Zhao, J. *et al.* IRESbase: a Comprehensive Database of Experimentally Validated Internal Ribosome Entry Sites. *bioRxiv* 2020.01.15.894592 (2020) doi:10.1101/2020.01.15.894592.

72. Mokrejš, M. *et al.* IRESite—a tool for the examination of viral and cellular internal ribosome entry sites. *Nucleic Acids Res.* **38**, D131–D136 (2010).

## Author contributions

Ya.Z., J.Z. and J.C. conceived and supervised the study. H.W. and Yi.Z. implemented the algorithms and performed the data analysis. Ya. Z and H. W. wrote the manuscript, and all authors read and contributed to the editing of the manuscript and approved the final version.

## Competing interests

All authors declare no financial interest. J.Z. is the founder and CEO and Ya.Z. is the scientific founder for Ribopeutic, respectively.

## Additional information

**Supplementary information** The online version contains supplementary material available at https://zenodo.org/records/14430869.

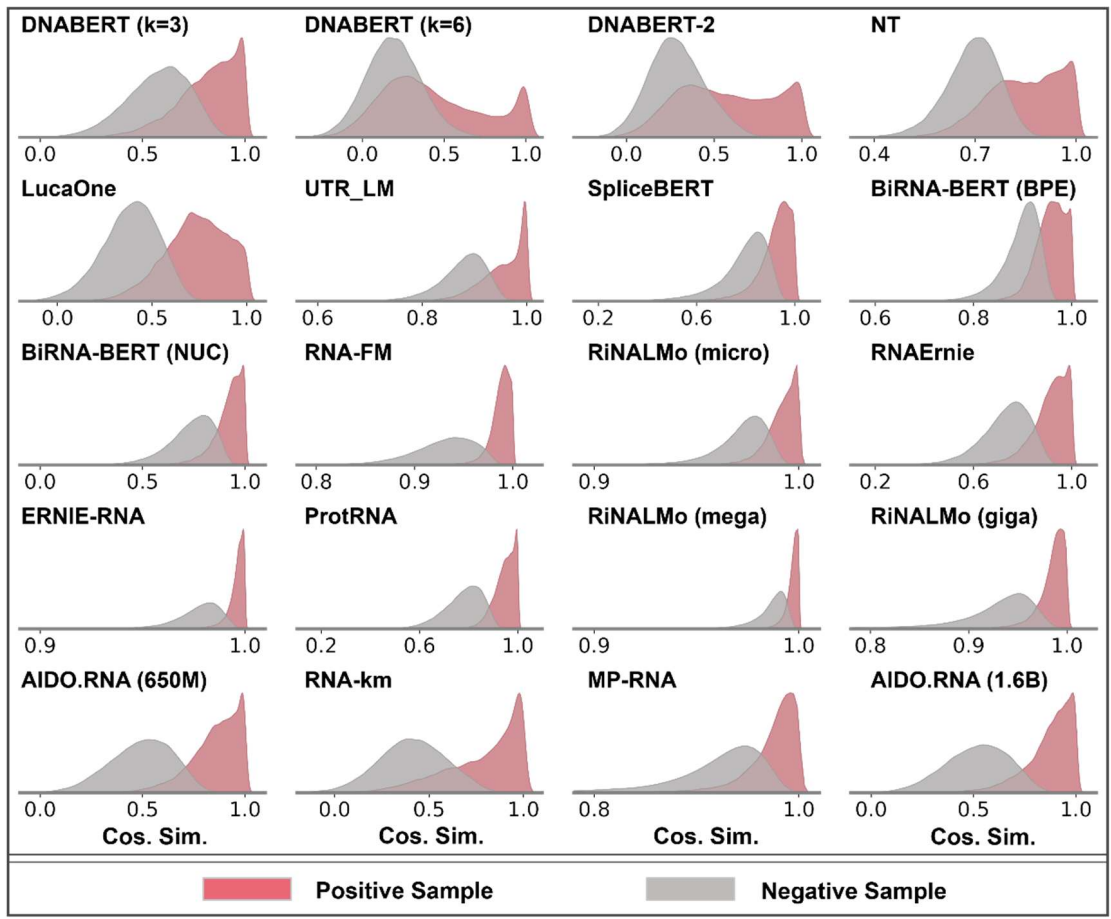

**Extended Data Fig. 1 | Cosine similarity distributions of sequence embeddings for RNAs in Rfam dataset.**

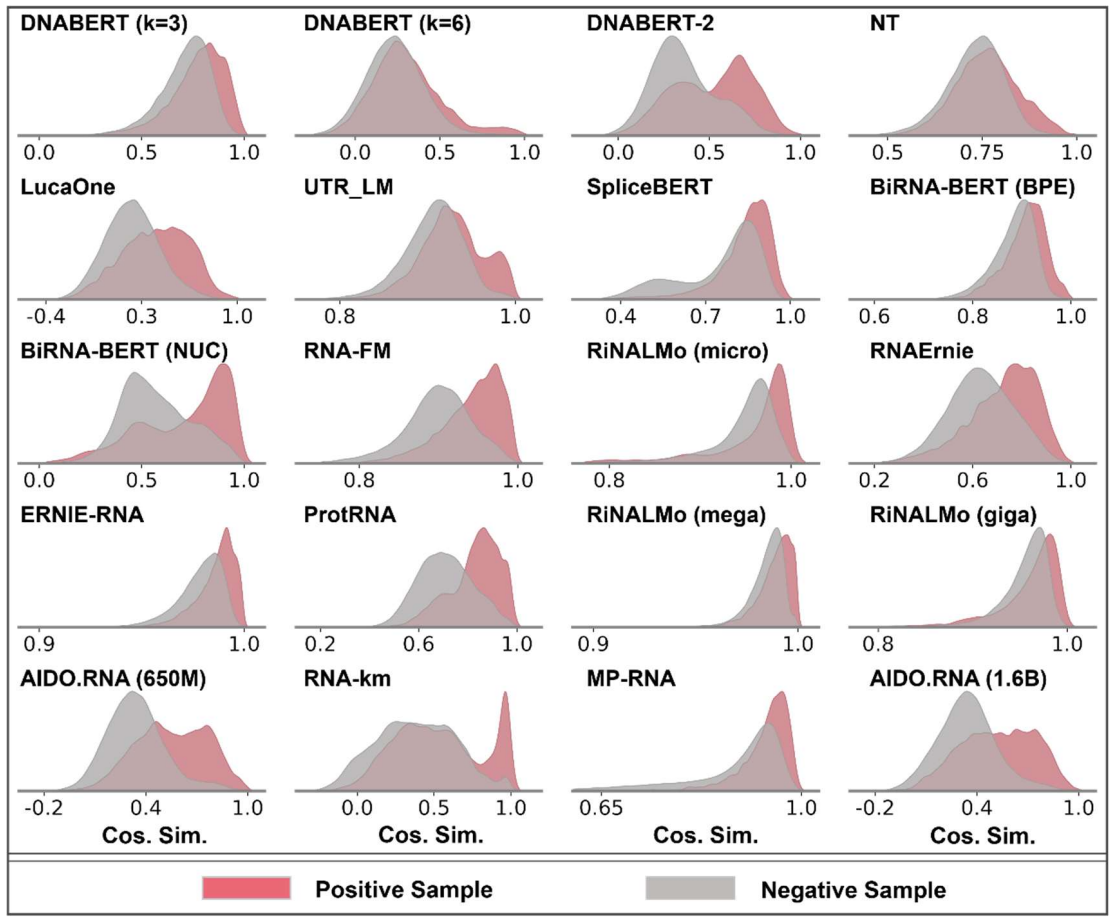

**Extended Data Fig. 2 | Cosine similarity distributions of sequence embeddings for RNAs in ArchiveII dataset.**

**Table 1.** A list of RNA language models with published models boldfaced[†].

| Class | Model | Year Posted | Ref. | Model Size | Pre-train Dataset | Input | Token | Network Art. | Num. of Layers | Num. of Heads | Embedding Dim | Max_Seq_Len | Remark |
|---|---|---|---|---|---|---|---|---|---|---|---|---|---|
| I | RNA-FM | 2022 | [17] | 100M | RNAcentral[29] (~23M Seq.) | Seq. | base | BERT | 12 | 20 | 640 | 1024 | - |
| | UNI-RNA[‡] | 2023 | [18] | ≤ 400M | RNAcentral & nt[54] & GWH[55] (~1B Seq.) | Seq. | base | BERT | 24 | 20 | 1280 | - | - |
| | **RNA-MSM** | 2024 | [20] | 96M | MSA (~4000 Rfam[32]) | MSA | base | MSA-Transformer | 10 | 12 | 768 | 1024 | - |
| | RiNALMo (micro) | | | 33M | | | | | 12 | 20 | 480 | | |
| | RiNALMo (mega) | 2024 | [24] | 148M | RNAcentral & nt & Rfam & Ensembl[56] (~36M Seq.) | Seq. | base | BERT | 30 | 20 | 640 | 1024 | - |
| | RiNALMo (giga) | | | 651M | | | | | 33 | 20 | 1280 | | |
| | ERNIE-RNA | 2024 | [19] | 86M | RNAcentral (~20.4M Seq.) | Seq. | base | BERT | 12 | 12 | 768 | 1024 | struct. enhanced |
| | **RNAErnie** | 2024 | [25] | 86M | RNAcentral (~23M Seq.) | Seq. | base | BERT | 12 | 12 | 768 | 512 | motif-aware mask |
| | RNA-km | 2024 | [26] | 152M | RNAcentral (~23M Seq.) | Seq. | base | BERT | 12 | 16 | 1024 | 512 | $k$-mers mask |
| | ProtRNA | 2024 | [23] | 651M | non-redundant RNAcentral (~6M Seq.) | Seq. | base | BERT | 33 | 20 | 1280 | 512 | transfer-learning |
| | BiRNA-BERT | 2024 | [34] | 116M | RNAcentral & RefSeq[54] (~36.5M Seq.) | Seq. | base (NUC) or BPE | BERT | 12 | 12 | 768 | ≥1024 (NUC) | dual-pretraining |
| | MP-RNA | 2024 | [30] | 186M | OneKP initiative[57] (~54.2B Bases) | Seq. & SS | base | BERT | 32 | 30 | 720 | 1024 | SS calibration |
| | **AIDO.RNA (650M)** | 2024 | [33] | 648M | RNAcentral (~41.5M Seq.) | Seq. | base | BERT | 33 | 20 | 1280 | 1024 | - |
| | **AIDO.RNA (1.6B)** | | | 1.6B | | | | | 32 | 32 | 2048 | 1024 | |
| II | **SpliceBERT** | 2024 | [27] | 20M | 72 vertebrate genomes from UCSC[58] (~2M Seq.) | Seq. | base | BERT | 6 | 16 | 512 | 1024 | - |
| | **UTR_LM** | 2024 | [21] | 1M | 5′ UTRs from Ensembl &Sample[39] & Cao[40] (~2.3M Seq.) | Seq. & SS & MFE | base | BERT | 6 | 16 | 128 | 1022 | - |

| | | | | | | | | | | | | |
|---|---|---|---|---|---|---|---|---|---|---|---|---|
| III | LucaOne | 2024 | [22] | 1.6B | Refseq (~1.3B Seq.) | Seq. | base | BERT | 20 | 40 | 2560 | 1280 | multi-level tasks |
| | Evo (131k) | 2024 | [28] | 6.5B | OpenGenome[28] (~316B Bases) | Seq. | base | StripedHyena | 32 | 32 | 4096 | 131,072 | - |

All models are sized as the number of parameters actually loaded during the model inference process, except for UNI-RNA whose model size is reported from the literature. For LucaOne, only the number of nucleic acid sequences in the pre-training are counted.

† "Seq.": RNA sequence; "SS": secondary structure; "MFE": the minimum free energy; "Input": input features during pre-training; "Max_Seq_Len": maximum input sequence length reported.

‡ The model is not open-source.

**Table 2.** Downstream tasks employed for model performance evaluation.

| Task | Type | LMs | Datasets |
|---|---|---|---|
| Structural: | | | |
| Solvent accessibility | Regression | RNA-MSM | Calculated from 3D RNA chain structures by the POPS package[59] with a probe radius of 1.4 Å |
| Secondary structure | Binary | RNA-FM, RNA-MSM, UNI-RNA, RiNALMo, ERNIE-RNA, RNAErnie, ProtRNA, MP-RNA, AIDO.RNA | RNAStralign[42], ArchiveII[43], bpRNA-1m[44], and datasets from SPOT-RNA and SPOT-RNA2 (TR0, VL0, TS0)[60,61] |
| 3D contact | Binary | RNA-FM, ERNIE-RNA | Datasets from RNAcontact[37], including non-redundant RNA 3D structures from Leontis and Zirbel[62] |
| 3D torsion angle | Regression | BiRNA-BERT | 3D RNA chain structures datasets from SPOT-RNA-1d[63] |
| Functional: | | | |
| Classification | Binary/Multi-class | BiRNA-BERT, RiNALMo, RNAErnie, LucaOne, AIDO.RNA | ArchiveII[43], Rfam[32] |
| Splice sites | Binary | SpliceBERT, RiNALMo, BiRNA-BERT, AIDO.RNA | Splice-site sequences datasets from Spliceator[64] |
| Modification sites | Multi-label | UNI-RNA, AIDO.RNA | 12 types of RNA modification sites from MultiRM[65] |
| RNA-protein binding | Binary | RNA-FM, ERNIE-RNA, ProtRNA, LucaOne | RNA-binding proteins in HeLa cell environment from PrismNet[38], ncRPI-LGAT[66] |
| RNA-RNA binding | Binary | RNAErnie, BiRNA-BERT | miRNA-mRNA interactions datasets from DeepMirTar[67]; miRNA-lncRNA interaction datasets from PmliPred[68] |
| Mean Ribosome Load | Regression | RNA-FM, UNI-RNA, RiNALMo, ERNIE-RNA, UTR_LM, ProtRNA, AIDO.RNA | 5'UTRs from Optimus 5-prime[39] |
| mRNA Ecp. Level and Trans. Efficiency | Regression | UTR_LM, AIDO.RNA | Gathered from human muscle tissue, human prostate cancer cell line PC3 and human embryonic kidney (HEK) 293T cell line[40] |
| IRES identification | Binary | UTR_LM | 46,774 mRNAs from multiple databases[32,69–72] |

Note: IRES: internal ribosome entry sites; RBPs: RNA binding proteins. The column "LMs" lists part of the RNA LMs that have been demonstrated to have applications in the corresponding downstream tasks; the column "Datasets" lists several frequently used datasets.

**Table 3**. Percentage of rRNA and tRNA in common ncRNA datasets.

| Dataset | # | rRNA | tRNA | other RNAs | Note |
|---|---|---|---|---|---|
| RNAcentral[29] | 32,785,565 | 61.0% | 22.5% | 16.5% | - |
| RNAStralign[42] | 37149 | 64.7% | 24.9% | 10.4% | - |
| ArchiveII[43] | 3,975 | 35.9% | 14.0% | 50.1% | - |
| bpRNA-1m[44] | 102,318 | 21.8% | 34.6% | 43.6% | - |
| TR0[60,61] | 10814 | 1.8% | 8.0% | 90.2% | - |
| Rfam 15.0 (seq.)[32] | 20,312,587 | 8.8% | 52.6% | 38.6% | 8.8% (miRNA) |
| Rfam 15.0 (fam.) | 4178 | 0.3% | 0.05% | 99.65% | 38.2% (miRNA) |

Note: For the Rfam dataset, the rRNA and tRNA percentages were counted in terms of number of sequences and number of families, respectively, and the miRNA with the largest percentage of families were noted.

**Table 4**. Zero-shot evaluation in structural properties (RNA secondary structure (or protein contact maps) and functional classifications (Rfam & ArchiveII) in term of median, mean, and std values for individual RNAs (proteins) using the same head-layer position for all RNAs (proteins)).

| Model | SS F1 (↑) | | Functional Classifications OR(↓) | |
|---|---|---|---|---|
| | Median | Mean± Std | Rfam | ArchiveII |
| **DNA LM** | | | | |
| DNABERT ($k$=3) | 0.006 | 0.017±0.023 | 0.255±0.001 | 0.627±0.003 |
| DNABERT ($k$=6) | 0.015 | 0.021±0.025 | 0.461±0.002 | 0.780±0.002 |
| DNABERT-2 | * | * | 0.363±0.000 | 0.493±0.002 |
| NT | * | * | 0.256±0.001 | 0.740±0.000 |
| **Class III (Unified LM)** | | | | |
| LucaOne | 0.175 | 0.223±0.195 | 0.154±0.000 | 0.498±0.002 |
| **Class II RNA LM** | | | | |
| UTR_LM | 0.022 | 0.029±0.032 | 0.202±0.001 | 0.626±0.000 |
| SpliceBERT | 0.034 | 0.043±0.039 | 0.240±0.001 | 0.578±0.001 |
| **Class I RNA LM** | | | | |
| BiRNA-BERT (BPE) | * | * | 0.206±0.001 | 0.626±0.003 |
| BiRNA-BERT (NUC) | * | * | 0.161±0.000 | 0.454±0.002 |
| RNA-FM | 0.180 | 0.267±0.239 | **0.070**±0.000 | 0.375±0.002 |
| RNAErnie | 0.238 | 0.288±0.223 | 0.156±0.000 | 0.514±0.000 |
| RiNALMo (micro) | 0.256 | 0.268±0.184 | 0.168±0.000 | 0.486±0.002 |
| ERNIE-RNA | 0.259 | 0.300±0.234 | 0.089±0.000 | 0.493±0.002 |
| ProtRNA | 0.298 | 0.333±0.236 | 0.081±0.000 | **0.361**±0.000 |
| RiNALMo (mega) | 0.315 | 0.342±0.225 | 0.146±0.000 | 0.587±0.001 |
| RiNALMo (giga) | 0.404 | 0.414±0.255 | 0.116±0.000 | 0.600±0.002 |
| AIDO.RNA (650M) | 0.438 | 0.402±0.234 | 0.104±0.000 | 0.395±0.001 |
| RNA-km | 0.473 | 0.492±0.236 | 0.197±0.001 | 0.683±0.001 |
| MP-RNA | 0.507 | 0.507±0.201 | 0.235±0.001 | 0.545±0.001 |
| AIDO.RNA (1.6B) | 0.579 | 0.508±0.263 | 0.110±0.000 | 0.445±0.000 |
| RNA-MSM | **0.631** | **0.582**±0.217 | * | * |
| **Protein LM** | | | | |
| ESM2 | 0.660 | 0.675±0.135 | * | * |

Note: "*" indicates that the task was not performed on the corresponding model. DNABERT ($k$=3) and DNABERT ($k$=6) indicate that every 3 bases and every 6 bases in the input sequence form a token when using DNABERT, respectively. Best values for RNA LMs in each column are bolded. Rfam contains a total of 24,607 sequences randomly sampled from 4170 families in Rfam 14.10 and ArchiveII is composed of 3,864 sequences from 9 RNA types.